\documentstyle[twocolumn,aps,prl,floats]{revtex}
\begin{document}
\date{Sub. to Chem. Phys. Lett 27-May 1994, Revised 1-July 1994 (to appear)}
\title{
Density-Functional Based Determination of the CH$_3$-CH$_4$ Hydrogen Exchange
Reaction Barrier}
\author{
Mark R. Pederson \\
Complex Systems Theory Branch \\
Naval Research Laboratory,
Washington D.C. 20375 \\ }
\maketitle
\tighten
\begin{abstract}
Due to the overbinding that is inherent in existing {\em local} approximations
to the density-functional formalism, certain reaction energies have not
been accessible. Since the generalized
gradient approximation significantly decreases the overbinding, prospects
for density-functional-based reaction dynamics are promising.
Results on the generalized-gradient based determination of the
CH$_{3}$-CH$_{4}$ hydrogen exchange reaction are presented.
Including all Born-Oppenheimer
effects an energy barrier of 9.5 kcal/Mole is found which is
a very significant improvement over the local-density approximation.
\end{abstract}
\section{Introduction}
 Over the past two or three decades quantum-mechanical methods
based on the Hohenberg-Kohn density-functional theory~\cite{lda}
have steadily evolved in scope,
complexity, numerical precision and intrinsic accuracy. A representative
albeit incomplete list of some of the
successes related to gaussian-orbital
density-functional applications
include the accurate prediction of molecular geometries~\cite{cvd111,andzelm}
and the accurate determination of molecular vibrational
frequencies.~\cite{andzelm,handy1,phonont,wang}
Specifically, in Ref.~\cite{cvd111}, the local density approximation (LDA)
was used to accurately calculate geometries of H$_2$, C$_2$, C$_2$H,
C$_2$H$_2$, C$_2$H$_4$,
C$_2$H$_6$, C$_6$H$_6$, CH$_3$C$_2$H and CH$_4$ and
in Ref.~\cite{andzelm}, Andzelm
{\em et al} showed that the
LDA would lead to accurate geometries {\em and} vibrational
modes for approximately ten other well
characterized molecules in the same
size regime.
Handy {\em et al}~\cite{handy1}
performed exceedingly
careful calculations on the benzene molecule and showed that
density-functional based
frequencies within {\em a local approximation} are particularily encouraging
since they are more accurate than the MP2 frequencies.  More recently
independent calculations
by Quong, Pederson and Feldman~\cite{phonont} and
Wang, Ho and Wang~\cite{wang} have shown that the local-density-approximation
reproduces the experimentally observed vibrational energies of the C$_{60}$
molecule to approximately 15-30 cm$^{-1}$.

Another interesting success
of the density-functional theory that is relevent to this paper is the work by
Salahub {\em et al}~\cite{salahub1,salahub2} and
more recently
Pederson {\em et al}~\cite{cvd111} and Handy {\em et al}~\cite{handy2}
on the methylene molecule. While the local-density-approximation underestimates
the
singlet-triplet energy splitting~\cite{cvd111,salahub1,handy2}, it leads
to the correct ordering and accurately describes the linear-bent energy
barrier of 5.3 kcal/mole~\cite{cvd111}.    Another example of an energy
barrier that is accurately determined within the LDA is the
eclipsed-staggered energy barrier of 2.8 kcal/mole~\cite{cvd111,dunlap} in
ethane.
These barriers are qualitatively similar since the bondlengths and
connectivities are essentially unperturbed during the distortion.
However, in a
reaction where bonds are broken and reformed, the overbinding in
LDA may seriously limit the accuracy of the reaction surface.
For example,
using a highly converged basis set,
consisting of a total of 65
even-tempered gaussian functions
on each carbon and hydrogen atom,
we have found that atomization energies
of nine hydrocarbon molecules are overestimated by approximately
0.7 eV per bond within the local density approximation.~\cite{perdewetal}
Since a typical reaction energy is
approximately 0.5-1.0 eV, the need for better accuracy is readily
apparent. Recently, Perdew and Wang~\cite{perdewt}
and Becke~\cite{becke} have developed
the generalized
gradient approximation (GGA) which has lead to vast improvements
in describing the bonding associated with light atoms
(at least).~\cite{cvd111,perdewetal,vanderbilt}
For example,
using geometries and densities from the local-density-approximation, we
find the average error in GGA bond energies to
be 0.1 eV for the hydrocarbon
molecules discussed in the first paragraph.~\cite{cvd111,perdewetal}
These energies have been compared by properly accounting for
effects due to zero-point energies.
Also, Salahub {\em et al}~\cite{salahub2}
and Handy {\em et al}~\cite{handy2} have shown that
the methylene singlet-triplet energy splitting is accurately
determined with gradient corrected energy functionals.
Due to the improvements realized from the generalized gradient
approximation,
the possibility for obtaining reaction energies within
a density-functional framework appears fruitful and some researchers
have begun looking at reactions  within the generalized gradient
approximation. For example,
Hammer {\em et al}~\cite{hammer} have studied the polarization and
charge transfer that occurs during H$_2$ dissociation on the aluminum
(110) surface.

Since the ability to calculate reaction surfaces quickly, precisely and
with chemical accuracy are prerequisite
to a first-principles
understanding of reactive processes and catalysis,
and since the density-functional formulation is
intrinsically faster than conventional quantum-chemistry shemes, this point
is investigated here.
The density-functional based
calculations presented here are numerically precise, converged with
respect to basis set, and include relaxation of the core electrons.
In this letter results on the
well-understood hydrogen exchange reaction
that occurs between a methyl [CH$_3$] radical and methane [CH$_4$]
molecule are discussed.
The experimental barrier is well characterized and is known to
be  about 14 kcal/mole.~\cite{expt}
This reaction barrier
has been studied
within several quantum-chemical techniques by several researchers
as well.~\cite{ther1,ther2,musgrave}

Many books on the subject of reaction dynamics exist~\cite{kinetics}
and
a recent discussion of how to calculate reaction dynamics from a
reactive surface is due to Truhlar and Gordon~\cite{truhlar}.
Here, the interest is in
determining how accurate a density-functional based
reaction surface may be and assessing the merits of the generalized
gradient approximation as compared to the local density approximation.
With respect to algorthmic issues,
even in the favorable case where intuition identifies
a small set (M) of active nuclear coordinates participating in
a chemical reaction, the quantum-mechanical determination of the
pathway over a reaction barrier requires approximately L$^M$ (with L=5-10)
times as many electronic-structure calculations than
are required to determine the energy of an N-atom system. Since scaling
with system size is already a problem for density-functional
electronic-structure-based
geometrical optimizations and {\em ab initio} quantum-chemical methods,
the need for intrinsic speed, chemical intuition and novel algorithmic
approaches to such problems is clear.  The
relative speed and more favorable scaling associated with
density-functional
based techniques is very attractive for these studies.

\section{Computational and Theoretical Details}
To perform the calculations discussed here,
the all-electron, full-potential
gaussian-orbital cluster code~\cite{codes1,codes1p,codes2,codes3} has
been used.
As
discussed in Ref.~\cite{codes1}, the potential is calculated analytically
on a variational integration mesh which allows for the
calculation of the electronic structure, total energies and Pulay-Corrected
Hellmann-Feynman~\cite{codes2,pulay}
forces with any desired numerical precision.
More recently,~\cite{phonont}
we have incorporated the calculation of
vibrational modes into the cluster codes. This is accomplished by
calculating the Hellmann-Feynman-Pulay forces at the equilibrium geometry
and neighboring points and using a
finite-differencing algorithm. Except where explicitly
stated,
moderately large contracted
gaussian-orbital basis sets, with gaussian decay parameters of
Huzinaga~\cite{huzinaga}, have been used. On each carbon atom, I
have used ten bare gaussians ($\alpha$=0.1146 to 4232.61)~\cite{huzinaga} to
construct a basis set of
8 s-type (3 of which have an r$^2$ prefactor), 4$\times$3=12 p-type, and
3$\times$5=15
d-type functions. On each hydrogen atom, I have used six bare
gaussians~\cite{huzinaga}
to construct a basis set of 3 s-type (1 of which have an r$^2$ prefactor),
1$\times$3=3 p-type, and 1$\times$5=5
d-type functions.
This amounts to a total of 133 single or contracted gaussian
orbitals on the seven atom complex.
This basis set has been developed over many years of
local-density-based simulations on hydrocarbon molecules, diamond, fullerene
molecules and tubules and is accurate.~\cite{cvd111}
As described in detail below,
we have further tested the
precision of these results by using a huge basis set of a total of 404
even-tempered gaussian functions distributed over the complex, and by using
this basis we determined
that the basis set discrepancies are exceedingly small.
The geometrical optimizations that were required for
these calculations utilized the conjugate-gradient algorithm and the
simulations proceeded until the forces were less than 0.01 eV/\AA. All
the results discussed here have been performed with the spin-polarized
density functionals.

\section{Results and Discussion}
Pictured in  Fig.~\ref{fig1} is a schematic diagram on how the hydrogen
exchange reaction proceeds. As discussed in the figure caption, the
reaction proceeds from configuration A to configuration C by allowing
the hydrogen atoms to relax as the two carbon atoms come close together.
The intrabond hydrogen overcomes a small barrier at the transition
state (Configuration B) and then relaxes downhill to configuration C.

Pictured in the lower panel of
Fig.~\ref{fig2} is the energy surface, as calculated
within the local-density-approximation (LDA). The minimum energy as a function
of C-C separation is shown in Fig.~\ref{fig3}. This corresponds to the energy
of the seven-atom complex as it travels along the lowest energy reaction
path in Fig.~\ref{fig2}.
As argued qualitatively above,
the LDA hydrogen-exchange reaction energy is significantly suppressed due
to the tendency to overestimate bond strengths. The final result is that
LDA leads to a classical energy barrier of approximately 0.7 kcal/mole which
is a factor of 20 smaller than experiment. The reactive surface has also
been determined within the framework of the generalized gradient
approximation of
Perdew {\em et al}.~\cite{perdewetal,perdewt} Using the techniques
discussed in Ref.~\cite{perdewetal}
the GGA reactive surface has been determined and is
shown in the upper panel of
Fig.~\ref{fig2}. The minimum LDA and GGA energies as a function of the
C-C separation (i.e. all hydrogenic degrees of freedom are relaxed) are
compared in Fig.~\ref{fig3}. Comparison of the reactive surfaces and pathways
show that the LDA and GGA lead to qualitatively different results.  Most
importantly, the GGA leads to a classical reaction barrier of  8.7 kcal/mole
which is reasonably close to the experimentally known barrier of
14 kcal/mole.~\cite{expt}
Also evident is that the LDA predicts a significantly
more compact reactive complex. Within LDA, the lowest geometry of the
CH$_4$-CH$_3$ complex is bound by 2.8 kcal/mole and has a C-C separation of
approximately 3.27 \AA.~ In contrast, within GGA, the lowest geometry is
only bound by 0.7 kcal/mole
and has a C-C separation of approximately 4.1 \AA.~
It is difficult to determine from experiment, whether the LDA or GGA
result is more accurate for this complex. However, in this regime,
the interactions are not
expected to be too different from the CH$_4$-CH$_4$ Van der Waals interactions
which are well understood.  For methane-methane interactions, the carbon
atoms are separated by 4.28 \AA~and are bound by 0.3 kcal/mole.~\cite{vdw}
For energies
of this size, the effects of zero-point motion must be taken into account
to accurately determine the enthalpy (rather than energy) of formation.
To determine the effects due to zero-point motion, the
vibrational frequencies of isolated CH$_3$ and CH$_4$ molecules and of
the bound CH$_3$-CH$_4$ complex have been calculated.
For methane, two triply degenerate vibrational
modes of 1231 and 3082 cm$^{-1}$, a doubly degenerate mode at 1473 cm$^{-1}$
and
a nondegenerate
mode at 2961 cm$^{-1}$ are found. These agree favorably with the experimental
modes of 1306, 3020, 1526 and 2914 cm$^{-1}$ respectively. For the
methyl-radical the vibrational modes within LDA are found to be
537, 3037 (nondegenerate), 1345, and 3158 (doubly dengerate) cm$^{-1}$.
By
subtracting the zero-point energies of
both the bound and isolated systems the enthalpy of formation
for the CH$_3$-CH$_4$ complex is found to be closer to 0.4 kcal/mole.
The fact that
the zero-point energy is larger for the complex than for the isolated
modes can be qualitatively explained by noting that for the isolated molecules
there are a total of twelve zero-energy modes.
However, when the two
systems complex, only six zero-energy modes
survive and six additional positive
energy modes emerge. Assuming that the original non-zero modes do not
significantly soften, an overall increase of the zero-point motion results.

Based on experience gained in many calculations on hydrocarbon molecules,
diamond and fullerene-based systems, the energetics obtained
with the original basis set are expected to be very accurate.
To further address this point, I
have used a huge set of gaussians to ensure the accuracy of these
calculations. On each carbon atom I have placed eleven even-tempered
s-gaussians
with $\alpha$=0.1 to 5000,
nine even-tempered
p-type gaussian sets with
$\alpha$=0.1 to 574.349, and four even-tempered d-type cartesian
gaussian
sets with $\alpha$=0.1 to 2.56857. On each hydrogen atom I have
placed seven
even-tempered s-type gaussians
with $\alpha$=0.08 to 138.889, five even-tempered p-type gaussian sets with
$\alpha$=0.08 to 11.556,  and three even-tempered d-type gaussian
sets with $\alpha$=0.08 to 0.9615. Since the cartesian polynomials in
front of d-type
gaussians can also lead to a spherical function of the form
r$^2$exp$(-\alpha r^2$), there are actually 15 and 10 spherical functions
on each carbon and hydrogen respectively. Taking account of the
degeneracies of the p and d sets this leads to 404 gaussians in the
basis set. The calculation of the energy along the reaction
path with this basis has
been repeated and the changes are indeed small with the
GGA and LDA barriers
increasing by 0.6 kcal/mole. It is noted that this {\em does not}
suggest that a larger basis necessarily increases the barrier and is
probably more indicative of the maximum change expected. For energy
scales this small, one should also consider that a small change in transition
state geometry could partially offset this small increase. Regardless, the
results of this paragraph show that the basis set used in these calculations
is sufficiently accurate to quantitatively determine the improvements that
are realized when a gradient-corrected functional is used in place of
a local functional.

As is well known, the simulation of a reaction requires the calculation of
a classical reaction surface and the additional ability to treat the
nuclei quantum-mechanically.~\cite{kinetics}
Methods for such simulations include
path-integral techniques or the techniques discussed in
Ref.~\cite{truhlar}. Here the interest is in qualitatively determining
the scale of the nuclear quantum mechanical effects and in assessing the
intrinsic numerical precision that is possible.
The vibrational modes have been calculated at the density-functional-based
transition state which is realized by
separating the two C-C atoms by
2.677 \AA,~ placing a
H atom midway between the two carbon atoms and then tying
off each of the C atoms by 3 hydrogens with C-H bond distances of
1.103 \AA~(configuration B in Fig.~\ref{fig1}).
Our geometry is in good agreement with the geometry
of Musgrave {\em et al.} who find a C-C separation of 2.72 \AA~ and a C-H
separation of 1.089 \AA.~~\cite{musgrave}
It is well
known that at this transition state
one expects six zero-energy modes, fourteen positive frequencies
and one imaginary frequency.
If the
translational and rotational modes are projected
out of the hessian matrix
prior to diagonalization,  the imaginary and
low-energy modes are found to be at
883i, 52, 280, 282, 507, 643 and 646 cm$^{-1}$. If the
rotational and translational degrees of freedom are not explicitly
removed prior to diagonalization, the zero point motion of
the complex changes by only 0.12 kcal/ mole but the results from the
more accurate projected
treatment of the hessian matrix are used here.

As noted above and pictured in Fig.~\ref{fig3}
the GGA energy of formation required to prepare this
state is 8.7-9.4 kcal/mole depending on whether the energy is measured
with respect to the separated  molecules (8.7) or the reactive complex (9.4).
Recently, Fan and
Ziegler and Deng, Fan and Ziegler~\cite{ziegler1,ziegler2}
have performed similar calculations on the
CH$_3$-CH$_4$ hydrogen exchange reaction within local and nonlocal
density-functional formalisms.  In good accord with the results of this
paper, they find a classical barrier of 11.7 kcal/mole with a gradient
corrected functional and 1.9 kcal/mole for a local functional.  The 1-2
kcal/mole deviation between their results and the results here is
significantly smaller than the deviations between different quantum-chemical
methods and should therefore be considered as good agreement since the
numerical methods used were quite different. Ziegler and Fan use
Slater-type orbitals, the frozen-core approximation, least-square-fits for
the solution of Poisson's equation, Vosko-Wilks-Nusair LDA,~\cite{vosko}
Perdew's nonlocal
correlation,~\cite{perdew86}
Becke's non-local exchange,~\cite{becke} and
Boerrigter's numerical integration~\cite{boerrigter}
scheme.  In contrast, in this work I use Gaussian-type orbitals,
relaxed cores, analytic solution of Poisson's equation,
Perdew-Zunger LDA,~\cite{perdewzunger}
the Perdew 91 non-local exchange and correlation~\cite{perdewt}
and the numerical
variational integration mesh of Pederson and Jackson.~\cite{codes1}
The calculations here and those of Ziegler {\em et al.}
use spin-polarized functionals throughout.
Given all the numerical
and formal differences, it appears that there is good agreement between
the philosophically identical density-functional based methods.

The reaction barrier for the CH$_3$-CH$_4$ hydrogen transfer reaction
has been studied with quantum chemical techniques as well and is known
to be about 14 kcal/mole experimentally.~\cite{expt} Several researchers
have used various quantum-chemistry algorithms to determine the
barrier and have obtained estimates in the range of
17-35 kcal/mole.~\cite{ther1,ther2,musgrave} While many more
calculations on other molecular systems are required before drawing
definitive conclusions, it appears that the generalized-gradient
approximation to the density-functional theory is
yielding results that
are as accurate as traditional quantum-chemical techniques for this
reaction. Further from the point of view of scaling with system
size, the density-functional framework is undoubtedly
more advantageous.  Hybrid methods, which employ fast density-functional
methods for pathway determination and more traditional calculations for
the final determination of energetics might also be useful.
\section{Acknowledgements}
  Thanks to Drs. J. Q. Broughton and J. Lill
for helpful discussions. This work was
supported in part by the Office of Naval Research. The calculations
were carried out on the CEWES CRAY C90 computer.

\newpage
\begin{figure}
\caption{Schematic diagram of reaction pathway followed during the
hydrogen exchange reaction. The lowest-energy pathway between
configuration A and C  is realized by
allowing the two carbon atoms to come within 2.677 \AA~from one another at
the transition state (configuration B). The remaining six hydrogen atoms
relax to accomodate the environment mandated by the C-C separation and
intrabond H location.
At the transition state configuration (B), there is one imaginery root
and the {\em classical}
equilibrium geometry is determined by allowing the C-C atoms
to separate and the active hydrogen to relax off center.
\label{fig1}}
\end{figure}
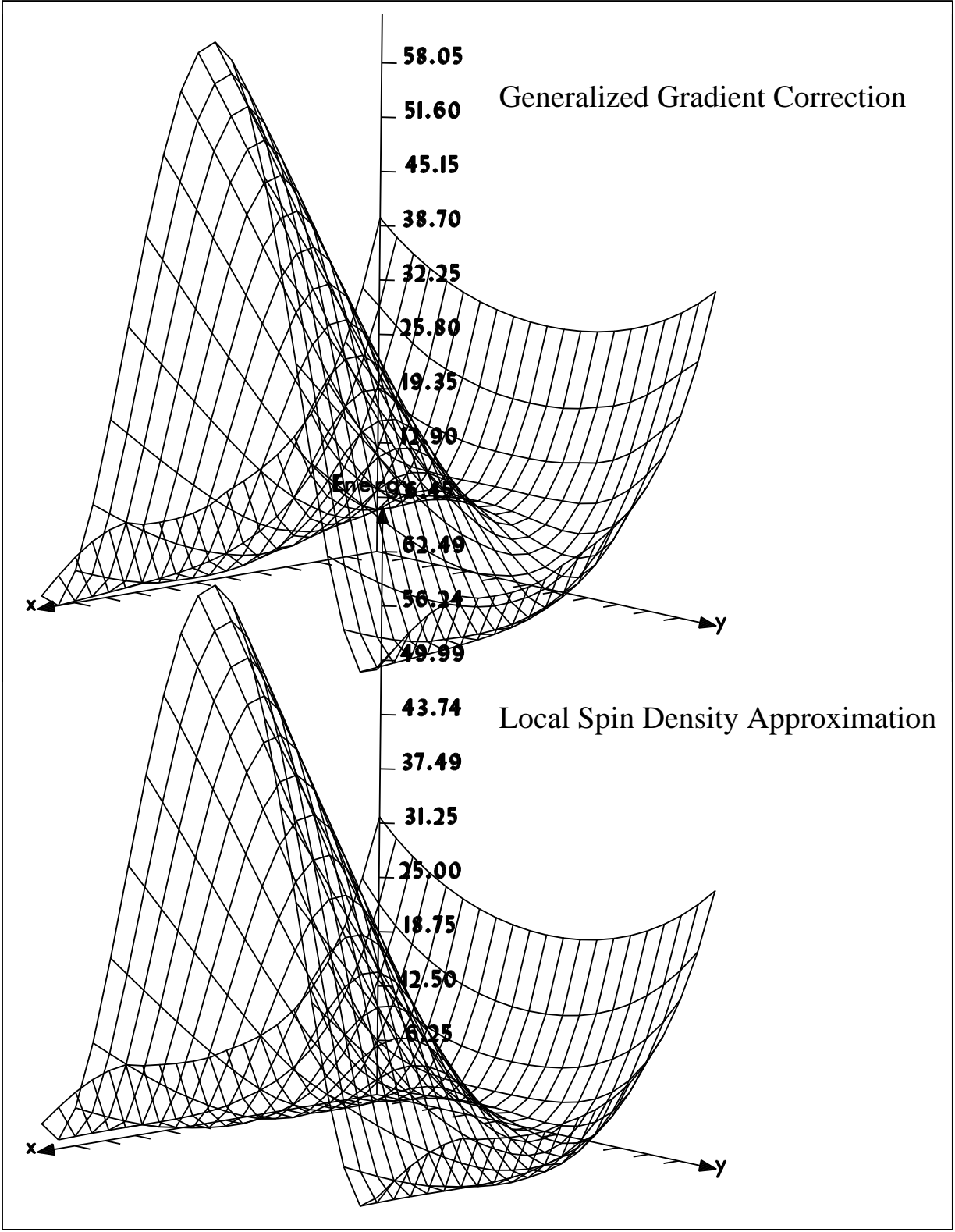
\begin{figure}
\caption{The LDA (a) and GGA (b) reaction surfaces. The y-axis corresponds
to the separation between the two carbon atoms and the x-axis corresponds
to the placement of the active hydrogen atom with respect to the bond
center. In contrast to LDA, which predicts a vanishingly small energy
barrier, the GGA predicts a classical barrier of 8.7-9.4 kcal/mole depending
on whether the energy is measured with respect to isolated or weakly
bound reactant states. Also
illustrated by the wire-mesh plots is the need to
determine the energy, interatomic forces and vibrational modes on
a two-dimensional Born-Oppenheimer surface of
variable convexity which enhances the computational cost  by
approximately 10$^2$ over a simple geometrical optimization
for the same system.
\label{fig2}}
\end{figure}
\begin{figure}
\caption{The minimum energy of the CH$_3$-CH$_4$ complex as a function of the
C-C separation. This potential curve was generated by allowing all hydrogenic
and electronic degrees of freedom to relax with the constrained C-C separation
and was derived by the lowest energy paths between the separated reactant
states and the transition state.  The geometries are schematically presented
at the top of the figure.
In comparison to GGA, the LDA predicts a very compact
and more strongly bound reactive complex. As is also shown in Fig.~2,
the energy of the transition state is significantly improved by the
addition of GGA.
The horizontal line at the minimum of the GGA potential represents
the excess zero-point motion that the complex
appropriates upon congregation. Including zero-point effects, the GGA
predicts that the complex is bound by approximately 0.4 kcal/mole. The
insert shows the energy along the entire reaction path.
\label{fig3}}
\end{figure}

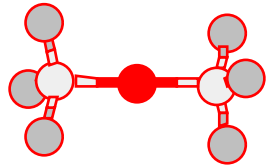
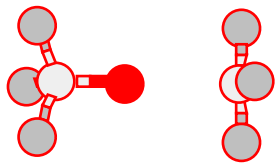
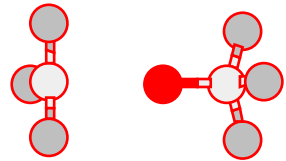
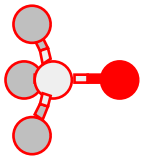
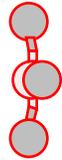
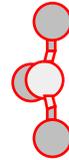
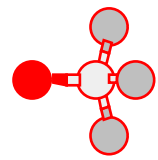
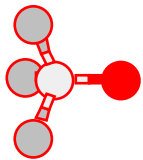
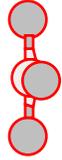
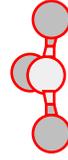
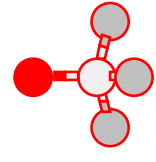
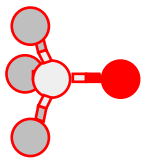
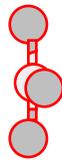
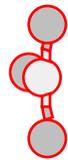
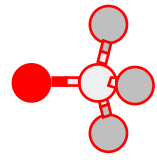
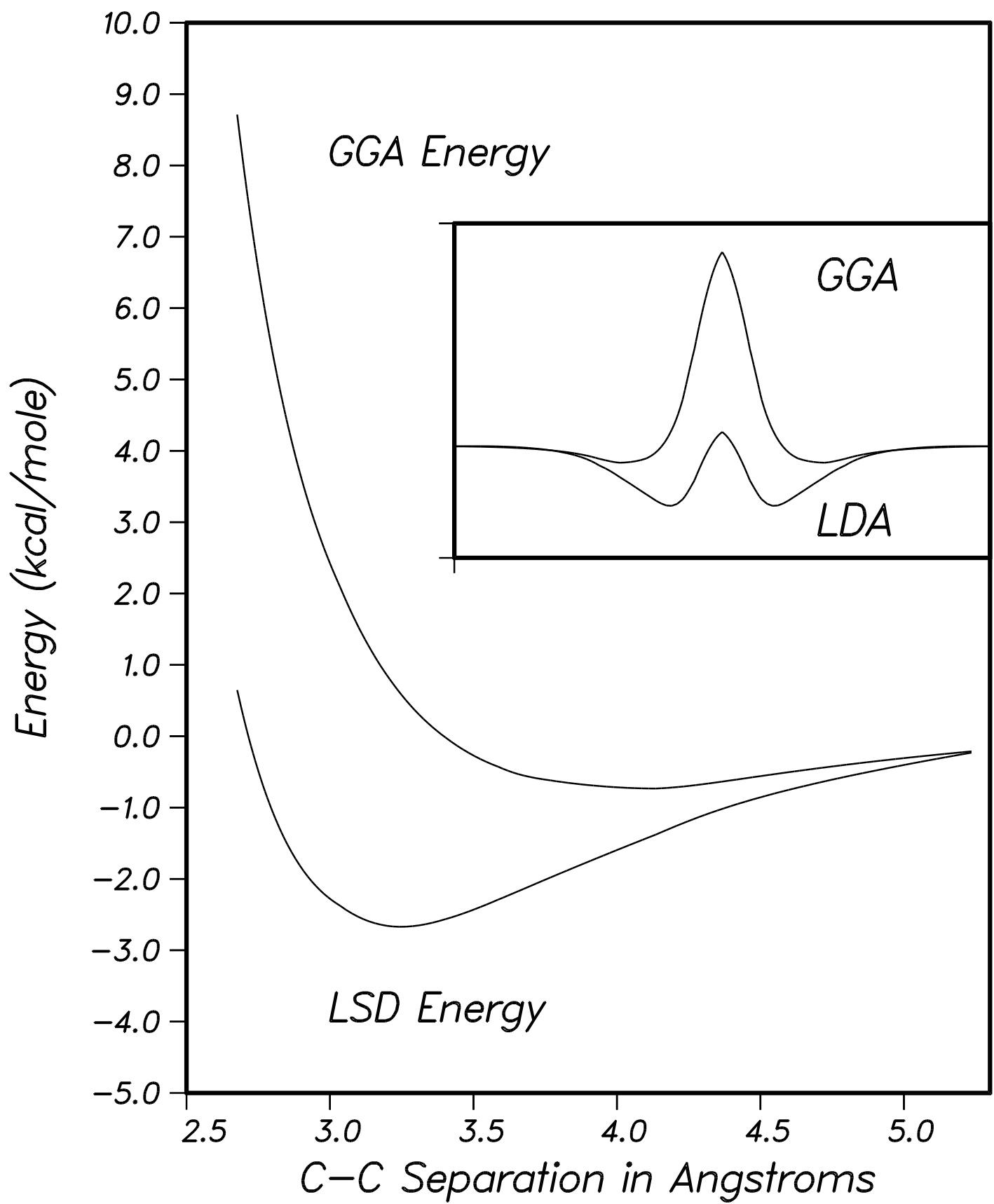
\end{document}